\RequirePackage{fix-cm}
\documentclass[twocolumn,epjc3]{svjour3}  
\smartqed  
\usepackage{amsfonts}
\usepackage{amssymb}
\usepackage{amsmath}
\usepackage{cite}
\usepackage{cancel}
\RequirePackage{graphicx}
\usepackage{dcolumn}
\usepackage{bm}
\usepackage{slashed}
\usepackage{lineno}
\usepackage{IEEEtrantools}
\usepackage{csquotes}
\usepackage[colorlinks = true,
            linkcolor = blue,
            urlcolor  = blue,
            citecolor = blue,
            anchorcolor = blue]{hyperref}
\usepackage{tikz,xcolor}
            
\DeclareOldFontCommand{\tt}{\normalfont\ttfamily}{\mathtt}
\definecolor{lime}{HTML}{A6CE39}
\DeclareRobustCommand{\orcidicon}{
	\begin{tikzpicture}
	\draw[lime, fill=lime] (0,0) 
	circle [radius=0.16] 
	node[white] {{\fontfamily{qag}\selectfont \tiny ID}};
	\draw[white, fill=white] (-0.0625,0.095) 
	circle [radius=0.007];
	\end{tikzpicture}
	\hspace{-2mm}
}

\foreach \x in {A, ..., Z}{\expandafter\xdef\csname orcid\x\endcsname{\noexpand\href{https://orcid.org/\csname orcidauthor\x\endcsname}
			{\noexpand\orcidicon}}
}

\journalname{Noname}
\begin{document}

\title{Leptons lurking in semi-visible jets at the LHC
}


\author{Cesare Cazzaniga \thanksref{e1,addr1}\orcidA{} 
        \and
        Annapaola de Cosa   \thanksref{e2,addr1}\orcidB{} 
}

\thankstext{e1}{e-mail: cesare.cazzaniga@cern.ch}
\thankstext{e2}{e-mail: adecosa@phys.ethz.ch}


\institute{ ETH Z\"urich Institute for Particle Physics and Astrophysics, CH-8093 Z\"urich, Switzerland \label{addr1}
}
\date{Received: date / Accepted: date}
\maketitle

\begin{abstract}
This Letter proposes a new search for confining dark sectors at the Large Hadron Collider. As a result of the strong dynamics in the hidden sector, dark matter could manifest in proton-proton collisions at the Large Hadron Collider in form of hadronic jets containing stable invisible bound states. These semi-visible jets  have been studied theoretically and experimentally in the fully hadronic signature where the unstable composite dark matter can only decay promptly back to Standard Model quarks. We present a simplified model based on two messenger fields separated by a large mass gap allowing dark bound states to decay into pairs of oppositely charged leptons. The resulting experimental signature is characterized by non-isolated lepton pairs inside semi-visible jets. We propose a search strategy independent from the underlying model assumptions targeting this new signature, and discuss the orthogonality with respect to the existing searches. Remaining agnostic on the shape of the di-lepton spectrum, we determine the sensitivity of a dedicated analysis to the target signal. The proposed search can claim the $3 \sigma$ evidence  (exclusion) of the heavier mediator up to masses of 3.5~TeV (4.5~TeV) with the full Run 2 data of the LHC. Exploiting the resonant feature of the lepton pairs can enhance the sensitivity reach on a specific model. We estimate that an analysis using the di-lepton invariant mass information can reach $5 \sigma$ discovery up to masses of 3.5~TeV and improve the exclusion up to more than 5~TeV.  
\end{abstract}

\section{\label{sec:intro} Introduction}
The Standard Model (SM) has collected many successes, culminating with the discovery of the Higgs boson~\cite{ATLAS:2012yve,CMS:2012qbp}. However, there are still open questions such as the origin of dark matter (DM), the neutrino masses and the baryon asymmetry, that cannot be addressed within the SM framework. 
The astrophysical and cosmological measurements of DM gravitational effects at large scales~\cite{Bertone:2004pz,Planck2018} strongly support the idea of a separate sector of particles whose spectrum has not been observed yet by the current searches~\cite{Liu:2017drf,Conrad:2017pms,Buchmueller:2017qhf}. \\
The most common strategy adopted by the LHC collaborations to search for DM relies on the assumption that DM is composed of Weakly Interacting Massive Particles (WIMPs)~\cite{PhysRevLett.39.165,Jungman:1995df}. WIMPs produce in the detector an excess of missing transverse momentum ($\cancel{E}_{\text{T}}$) accompanied by other visible objects, e.g. jets, photons, leptons~\cite{CMS:2021far,ATLAS:2021kxv,CMS:2018ffd,CMS:2022yjm,ATLAS:2020yzc,CMS:2019zzl,CMS:2020ulv,ATLAS:2014wfc} or by a Higgs boson~\cite{CMS:2019ykj,ATLAS:2021shl} ($\cancel{E}_{\text{T}}+$X signatures).\\
 Hidden Valley models~\cite{Strassler:2006im} propose alternative, well-motivated BSM scenarios to explain DM nature. These models arise in many top-down approaches, including String theory~\cite{Cvetic:2002qa,Arkani-Hamed:2005zuc} and they appear consistent with many UV-complete theories aiming to solve fundamental problems such as the electroweak hierarchy~\cite{PhysRevLett.96.231802,Burdman_2007,Craig:2015pha}. Hidden Valley models with strongly coupled hidden sectors can evade the phase-space probed by the collider searches mentioned above. In this scenario, novel experimental signatures emerge, characterized by sprays of particles resembling hadronic jets that include DM bound states and their decay products (dark jets).  Dark jets can appear in the detector at different distances according to the lifetime of the hidden sector bound states. In case of long-lived DM bound states,  displaced signatures arise, such as those characterizing emerging jets~\cite{Schwaller:2015gea} and trackless/displaced jets~\cite{ATLAS:2019qrr}. For prompt decays of the dark sector bound states to SM particles, typical signatures are prompt dark jets~\cite{Park:2017rfb,Cohen:2020afv} or semi-visible jets (SVJ)~\cite{Cohen:2015toa,Cohen:2017pzm,Bernreuther:2019pfb}. In particular, semi-visible jets occur if a sizable fraction of the DM bound states within the dark jets remains stable, resulting in a multijet+$\cancel{E}_{\text{T}}$ signature characterized by $\cancel{E}_{\text{T}}$ aligned with one of the jets. To evade the strong experimental constraints from high mass di-lepton searches~\cite{CMS:2018ipm,ATLAS:2019erb}, available  semi-visible jets simplified models assume a TeV-scale leptophobic $Z'$ boson as unique mediator between the dark sector and the SM. Under this hypothesis, leptons cannot be produced directly from the decays of the DM bound states mediated by the messenger boson. This assumption poses some limits in terms of possible experimental signatures and DM discovery perspectives. In this Letter, we present a simplified model developed on concepts proposed in~\cite{Knapen:2021eip} leading to semi-visible jets enriched in non-isolated leptons (SVJ$\ell$), and propose a general search strategy to probe this novel signature at the LHC experiments. 

\section{\label{sec:model}Simplified model setup}

Driven by our target experimental signature, characterized by prompt non-isolated leptons produced inside semi-visible jets, we introduce a simple Hidden Valley model that will enable us to estimate the LHC sensitivity to this class of signals. The model is built in the Simplified DM Models fashion~\cite{ABDALLAH20158} and it is presented for illustration purposes since the experimental signature we are targeting in this study is common to  a large class of dark-sector theories. \\ In the Hidden Valley scenario that we consider, the SM gauge group is supplemented by a non-abelian dark sector $SU(N_c)_v$ with gauge coupling $g_v$. Here we choose  the number of dark color charges $N_c =3$ for concreteness. Within this hidden sector, the only fermions in the fundamental representation of $SU(3)_v$ are the dark quarks $q_{v,i}$, with $i \in \{1, \cdots , N_f \}$, where $N_f$ is the number of dark flavors. Confinement of this Yang Mills theory at a scale $\Lambda_v$ is guaranteed only for $N_f < 3N_c$~\cite{Albouy:2022cin}. In our study, we assume a minimal two dark flavors model ($N_f = 2$) with mass degenerate $q_{v,i}$ states. The dark sector is assumed not to be completely secluded from the SM. Following the same approach as in~\cite{Cohen:2015toa,Cohen:2017pzm}, we require a first TeV-scale $U(1)'_v$ leptophobic messenger field $Z^{'}$ to couple directly with both the SM quarks axial-vector current $J_{q,SM}^{\mu}$ and the dark quarks vector current $J_{v}^{\mu}$~\cite{Backovic:2015soa}.
\begin{figure}
\includegraphics[scale=0.32]{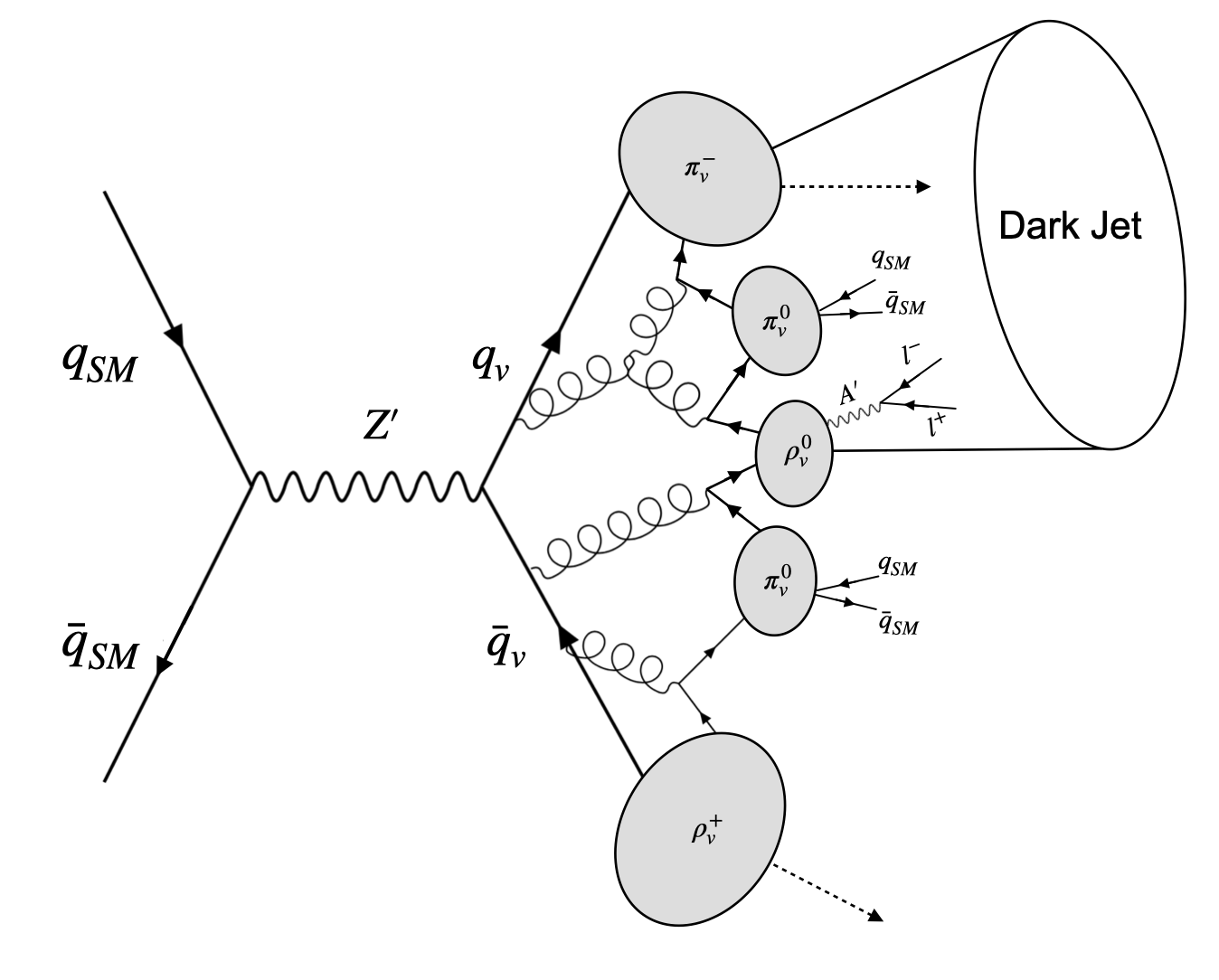}
\caption{\label{fig:s-channel SVJL diagram}S-channel production of semi-visible jets with non-isolated prompt leptons produced from dark hadrons decays.}
\end{figure}
The $Z'$ boson can acquire its mass $M_{Z^{'}}$ via spontaneous symmetry breaking of the $U(1)^{'}_v$ due to a dark Higgs field with non-zero vacuum expectation value~\cite{Strassler:2006im}. The dark Higgs sector has been neglected as it is assumed to be out of the LHC reach and therefore irrelevant for the phenomenology discussed here.\\ The energy scale of the event is set by \mbox{$M_{Z^{'}} \sim 1-6 \ \text{TeV}$}. Requiring the coupling of $Z'$ to dark quarks ($g^{v}_{Z^{'}}$) to be much larger than the coupling to SM quarks ($g^{q}_{Z^{'}}$), the $Z'$ boson is expected to decay predominantly into dark quarks $q_{v,i}$. A QCD-like parton shower in the dark sector initiated by $q_{v,i}$ is considered~\cite{HOOFT1974461,Knapen:2021eip}. The soft and collinear splittings draw the energy scale of the process down to the dark confinement scale \mbox{$1 \ \text{GeV} < \Lambda_v < 1 \ \text{TeV}$}  where dark hadrons of spin 0 (pseudo-scalar mesons $\pi_v$) and spin 1 (vector mesons $\rho_v$) are produced after hadronization. Some of these dark hadrons will be stable, and others will decay back to SM according to the conservation of global accidental symmetries of the dark sector, such as dark-isospin and dark-baryon number~\cite{Strassler:2006im,Cohen:2015toa,Cohen:2017pzm}. To allow DM bound states to decay to leptons, we follow a similar approach as in~\cite{Knapen:2021eip}, and we introduce a light $U(1)^{''}_v$ vector messenger field $A_{\mu}^{'}$. The $A_{\mu}^{'}$ field mixes with the SM hypercharge field $B_{\mu}$ via a term $\epsilon/2 \ F^{\mu \nu}[A']F_{\mu \nu}[B]$ governed by the parameter $\epsilon$.  After the diagonalization of the mixing term~\cite{PhysRevD.75.115001}, the Lagrangian $\mathcal{L}^{''}$ for the $A^{'}$ messenger sector is:
\begin{eqnarray}\label{eq:lagr2} 
\mathcal{L}^{''} \supset &  -\frac{1}{4} F_{ \mu \nu}[A'] F^{\mu \nu}[A'] + \frac{1}{2}M^2_{A^{'}} A^{'}_{\mu} A^{'\mu} - \epsilon e Q A^{'}_{\mu} J_{q,SM}^{\mu} \nonumber \\& - \epsilon e Q A^{'}_{\mu} J_{l,SM}^{\mu}  - g^{v}_{A^{'}} A^{'}_{\mu} J_{v}^{\mu}.
\end{eqnarray}
The mass $M_{A'}$ of the $A'$ boson can be generated in the same way as the $Z'$ mass. However, we assume $A'$ to couple much more weakly to the dark Higgs field compared to the $Z'$ boson, therefore leading to \mbox{$M_{Z'} \gg M_{A'}$}, with $A'$ at the GeV-scale. This mass hierarchy favors the decay of the unstable dark vector mesons via the lighter messenger $A'$. Indeed, the decay width of these DM bound states scales with the mediator mass as $\sim 1/M_{\text{med}}^4$, thus suppressing the off-shell $Z'$-mediated decays. The dark pseudo-scalar mesons $\pi_{v}$ can decay promptly through the $Z'$ portal to heavy quarks via a mass insertion \cite{Cohen:2015toa} due to helicity suppression.  \\
The partial widths of an unstable dark vector meson $\rho_v$ decaying to SM fermions via the $A'$ boson can be calculated from a chiral EFT~\cite{Knapen:2021eip}.
The parameter governing the mixing in the chiral EFT is given by $\epsilon_{\text{eff,v}}$ incorporating the non-perturbative dark sector dynamics information together with the gauge mixing parameter~$\epsilon$, and an effective mass scale dependence $ \sim 1/ M_{A'}^4$.  \\
The immediate phenomenological consequence of the two messenger fields model is a factorization between the production of the dark quarks, which can be mediated by a heavy leptophobic $Z'$ boson, and the decays of the dark vector mesons, which are completely driven by the lighter $A'$. As sketched in Figure \ref{fig:s-channel SVJL diagram}, this model leads to semi-visible jets with inside pairs of opposite charge leptons produced by the unstable dark hadrons. A further production mechanism for dark quarks could be directly via the $A'$ boson. However, this production channel contributes mainly to low mass SVJ$\ell$, where current trigger selection strategies dramatically limit a search sensitivity. For this reason, we don't investigate this contribution here and leave it to future studies. Lower bounds for the life-times $c \tau_{\text{min}}$ of the unstable dark vector mesons have been calculated in~\cite{Knapen:2021eip}. For DM bound states with masses at least around the GeV-scale, $c \tau_{min}$ can be so small that detectable displacements
between the production and decay points of unstable
dark hadrons cannot be resolved,  resulting in a signature with prompt leptons.

\section{Monte Carlo Simulations}

The signal and background processes are produced at $\sqrt{s} = 13$ TeV with
{\tt MadGraph5\_aMC@NLO}~\cite{Alwall:2014hca} event generator using the parton distribution functions {\tt NN23LO1} ~\cite{Ball:2013hta} from the {\tt Lhapdf}~\cite{Buckley:2014ana} repository. The evolution of the parton-level events and hadronization are performed with {\tt Pythia8}~\cite{Bierlich:2022pfr}. The initial state partons are matched with the jets using the MLM scheme~\cite{Alwall:2007fs} implemented in {\tt MadGraph5}. Initial and final-state radiation as well as underlying events are included in the simulation. To simulate the showering, hadronization and decays in the dark sector, we used the Hidden Valley module~\cite{Carloni:2011kk,Carloni:2010tw} in {\tt Pythia8}. The generated events are then interfaced with {\tt Delphes3}~\cite{deFavereau:2013fsa} to model the response of the CMS detector~\cite{CMS:2008xjf}. In our study, anti-$k_T$ jets~\cite{Cacciari:2008gp,Cacciari:2011ma} with \mbox{$R = 0.8$} (AK8 jets) are reconstructed requiring a minimum $p_{\text{T}}$ for clustering of 200 GeV. We  use AK8~jets since semi-visible jets are expected to be wider than typical SM jets because they arise from a multi-step process beginning with dark quarks produced via $p p \to Z' \to q_v \bar{q}_v$ and ending up with SM hadrons after hadronization in the hidden sector and in the SM~\cite{CMS:2021dzg}. All the samples have been normalized to the LHC Run~2 integrated luminosity $\mathcal{L}_{int} = 138 \ {\text{fb}^{-1}}$. \\ \\
For the SVJ${\ell}$ s-channel signal process, $20 \cdot 10^3$ events have been generated at LO for \mbox{$M_{Z'} \in [1.5, 5]$ TeV} with 500~GeV steps. Following the most recent recommendations from the LHC
DM Working Group~\cite{Boveia:2016mrp}, the coupling $g^{q}_{Z'}$ has been assumed to be flavour-universal and set to 0.25. Taking into account the number of flavours and colours in the dark sector as in~\cite{CMS:2021dzg}, the coupling $g^{v}_{Z'}$ has been fixed to 0.4. For this set of couplings, the constraints from the di-jet searches are still weak~\cite{CMS:2021dzg,CMS:2019gwf,Cohen:2015toa}. The number of stable and unstable dark hadrons produced in the dark hadronization process can vary according to the details of the dark sector. To capture these variations, an effective invisible fraction can be defined as $r_{\text{inv}}=  \langle {N_{\text{stable}}}/{(N_{\text{stable}} + N_{\text{unstable}})} \rangle$.
This parameter can take any value between 0 and 1~\cite{Cohen:2015toa,Cohen:2017pzm} depending on the details of the dark sector such as the number of dark flavours, the number of dark colors, dark quark masses, fragmentation and accidental global symmetries. To span a wide range of possible experimental signatures as a function of the model parameters~\ref{table:signal parameterization}, we consider three  different benchmark scenarios for $r_{\text{inv}}$ representing different possible regimes.  
The effect of $r_{\text{inv}}$ is simulated by requiring dark hadrons to decay to a pair of invisible particles with probability~$r_{\text{inv}}$. The dark hadrons that decay invisibly are proxies for the stable dark hadrons, while the remaining dark hadrons are the unstable ones. \\
Even for mass degenerate dark quarks, the dark pseudo-scalar $\pi_v$ and dark vector $\rho_v$ mesons masses can differ according to the hidden sector non-perturbative dynamics. Setting $\Lambda_v$ as overall mass scale for dark mesons, lattice QCD fits in~\cite{Albouy:2022cin} have been used to predict the masses of dark vector mesons $m_{\rho_v}$ from the input ratio  $m_{\pi_v}/\Lambda_v$. As a benchmark, the hadronization scale $\Lambda_v$ has been set to 5 GeV and the ratio $m_{\pi_v}/\Lambda_v = 1.6$, therefore fixing $m_{\pi}$ in the heavy quarkonium region at 8 GeV and $m_{\rho_v} \sim $ 15 GeV. 
The effective mixing parameter $\epsilon_{\text{eff,v}}$ has been set saturating the electroweak precision tests bound~\cite{Curtin:2014cca,Knapen:2021eip} evading the existent constraints from searches for dilepton resonances~\cite{CMS:2018ipm,ATLAS:2019erb,PhysRevLett.124.041801}. With this choice of the input parameters, the predicted average fraction of unstable $\rho_v$ mesons decaying into leptons is $\sim 15 \%$ per each flavour. The life-times of the unstable dark hadrons $c \tau_{\text{min}}$ has been fixed such that no displacement can be resolved experimentally according to theoretical lower bounds for dark hadron masses at the GeV-scale. Table~\ref{table:signal parameterization} summarizes the five most relevant parameters for signal modelling. Three are sensitive to the details of the dark sector: the confinement scale $\Lambda_v$, the pseudo-scalar mass ratio $m_{\pi_v}/\Lambda_v$ and the invisible fraction $r_{\text{inv}}$. Moreover, there are two portal parameters: the mass of the $Z'$ boson $M_{Z'}$, and the effective mixing  $\epsilon_{\text{eff,v}}$. \\
\begin{table}[!htbp]
\renewcommand{\arraystretch}{2}
\centering
\sffamily
\begin{tabular}{ccc}
\small PARAMETER  & \small DESCRIPTION & \small BENCHMARK \\ \hline
 $M_{Z'}$ & $Z'$ pole mass  & 1.5-5 TeV   \\  
 $\epsilon_{\text{eff,v}}$ & effective mixing  &  0.03 \\ 
$r_{\text{inv}}$ & invisible fraction  & 0.3, 0.5, 0.7    \\ 
$\Lambda_v$ & dark confinement scale  & 5 GeV  \\  
$m_{\pi_v}/\Lambda_v$ & pseudo-scalar mass ratio  &  1.6 \\ \hline 
\end{tabular}
\caption{Signal model parameterization.}
\label{table:signal parameterization}
\end{table}\\
\renewcommand{\arraystretch}{1}
\noindent
In this study, we have considered the same background processes as in the CMS semi-visible jets search~\cite{CMS:2021dzg}. All the background samples have been generated at LO. The QCD sample ($4 \cdot 10^7$ events) has been produced requiring a generator level cut on the leading parton jet $p_{\text{T}} > 500$ GeV. The QCD background is particularly relevant due to the large cross-section and the possibility of mis-reconstruction of the jet momentum leading to additional missing momentum aligned with the jet axis. The presence of b-flavored hadrons decaying into leptons inside the jets further mimics the signal signature. The $\text{t} \bar{\text{t}} + \text{jets}$ inclusive sample ($5 \cdot 10^7$ events) has been generated with up to two additional partons. This background mainly becomes of relevance when the top quarks are boosted, and therefore the W boson decay and the b-initiated jet are merged into a larger jet. The electroweak inclusive backgrounds $\text{Z}(\nu \bar{\nu}) + \text{jets} $ and $\text{W} (\ell \nu) + \text{jets}$ (both $2.5 \cdot 10^7$ events) have been produced with a generator level cut $H_{\text{T}} > 100$ GeV and including up to three additional partons in the matrix element. 

\section{Search strategy }

\begin{figure*}
\centering
\includegraphics[width=0.74\textwidth]{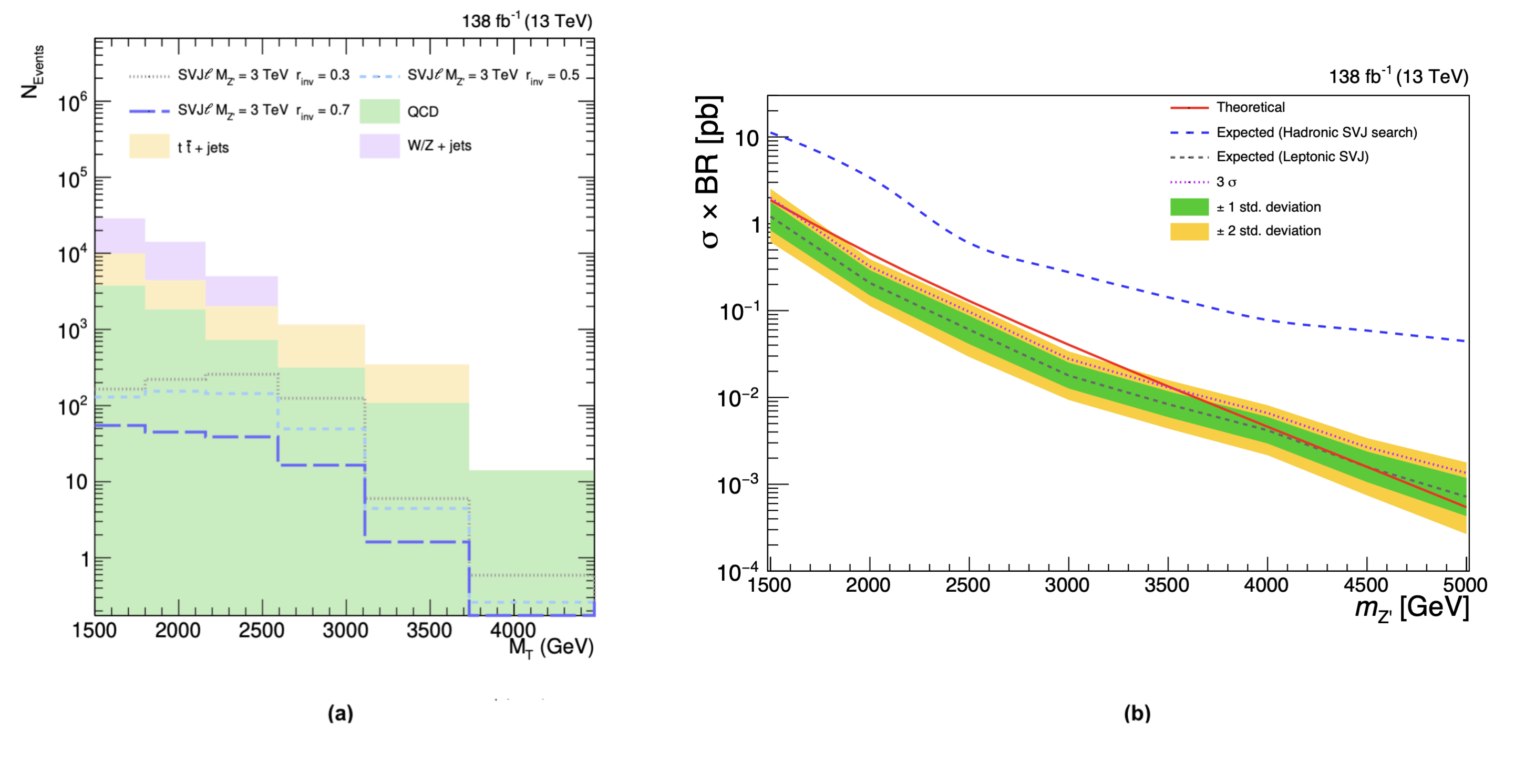}
{\caption{\textbf{(a)} $M_{\text{T}}$ template for the signal (benchmark mass point $M_{Z'}= 3$ TeV and  $r_{\text{inv}} = 0.3, \ 0.5, \ 0.7$) and background processes after applying the selections proposed in this Letter (without the requirement on $m_{\ell \ell}$). The binning has been chosen according to an experimental resolution $ \sim 20 \%$.  \textbf{(b)} Expected  limits on $\sigma \times \text{Br}(Z' \to q_v \bar{q}_v)$ for the signal benchmark with $r_{\text{inv}}= 0.3$ and all the other parameters fixed as in Table \ref{table:signal parameterization}. The fully hadronic SVJ analysis expected exclusion reach (dashed blue line) for the SVJ$\ell$ signature is compared with the proposed SVJ$\ell$ cut based analysis without exploiting the selection on $m_{\ell \ell}$ (dashed gray line).  \label{fig:Limits}}}
\end{figure*}

The main feature of the SVJ${\ell}$ signal is the presence of pairs of leptons from dark hadrons decays. The hadronic activity near a reconstructed object such as a lepton $\mathcal{\ell}$, is quantified by the  isolation variable  $\text{I}({\mathcal{\ell}})$~\cite{deFavereau:2013fsa}. The isolation is defined as the sum of the momenta of all the  particles reconstructed within a cone of given radius centred around the lepton itself,  normalized to lepton $p_{\text{T}}$. The isolated leptons ($\text{I}({\mathcal{\ell}}) < 0.1$) have a small probability to originate from a jet. Due to a large number of constituents within a jet cone, SVJ${\ell}$ leptons pairs fail the standard isolation requirements applied in BSM searches~\cite{CMS:2012baj,ATLAS:2014gmw,CMS:2018ncu,CMS:2019buh,ATLAS:2014vih,CMS:2018ipm,ATLAS:2019erb,PhysRevLett.124.041801}.
\begin{figure*}[!htbp]
\centering
\includegraphics[width=0.9\textwidth]{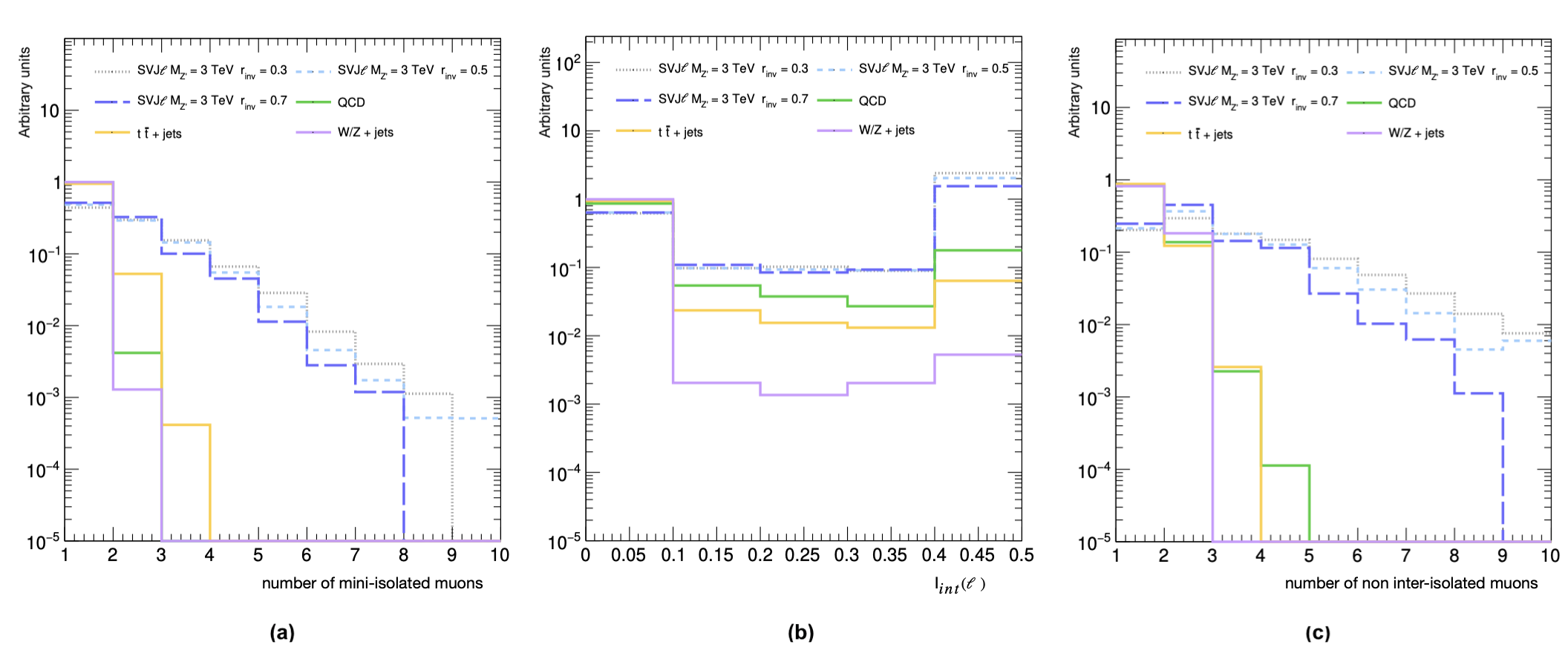}
{\caption{\textbf{(a)} Multiplicity of mini-isolated muons. \textbf{(b)} Distribution of muons inter-isolation for $R_{\text{iso}}= 0.5$. \textbf{(c)} Multiplicity of non inter-isolated muons. In all the histograms the last bin represents the overflow. The signal distributions are referred to a benchmark mass point $M_{Z'}= 3$ TeV and $r_{\text{inv}} = 0.3, \ 0.5, \ 0.7$, while all the other parameters are fixed as in Table \ref{table:signal parameterization}. \label{fig:leptons multiplicities}}}
\end{figure*}
\noindent
Furthermore, in the fully hadronic SVJ search~\cite{CMS:2021dzg}, the background from boosted top decays in semi-leptonic final states is reduced by applying a veto on mini-isolated leptons~\cite{Rehermann:2010vq,CMS:2020cur}.
However, this veto is highly inefficient to select SVJ${\ell}$ events, because some leptons inside SVJ${\ell}$ are expected to be mini-isolated as shown in Figure \ref{fig:leptons multiplicities}\textcolor{blue}{.a}. \\
To exploit the lepton-enriched content of SVJ${\ell}$ signal and enhance background rejection, we define an inter-isolation variable $\text{I}_{\text{int}}(\ell) = {\sum_{\Delta R} p_{\text{T},\ell_j}}/ {{p_{\text{T},\ell}}}$ quantifying how much each lepton $\ell$ is isolated with respect to all the other leptons $\ell_j$ within a cone of radius $R_{\text{iso}}$.
As~shown in Figure \ref{fig:leptons multiplicities}\textcolor{blue}{.b}, inter-isolation provides an evident handle against all the backgrounds and it remains almost independent from $r_{\text{inv}}$. Furthermore, this variable captures the main feature of the SVJ${\ell}$ leptons leading to a peak in the multiplicity for one pair of non inter-isolated leptons as evidenced in Figure \ref{fig:leptons multiplicities}\textcolor{blue}{.c}. Due to its features, inter-isolation can be used as an additional requirement to select SVJ${\ell}$ candidates.\\
In this letter, we propose a cut-based search strategy for leptons-enriched semi-visible jets. We apply  the selections used in the inclusive fully hadronic SVJ search~\cite{CMS:2021dzg}. For completeness, we explicit the selections in the following paragraphs. On top of it, we introduce a  set of model-independent requirements exploiting the leptonic content of SVJ${\ell}$. \\
We require each event to have at least two AK8 jets with $p_{\text{T}} > 200$ GeV and $|\eta| < 2.4$. Furthermore, the two leading jets are required to have a maximum pseudo-rapidity separation $\Delta \eta(j_1, j_2) < 1.5$, removing the large t-channel QCD contribution~\cite{CMS:2013egk,ATLAS:2014ktg}. \\
The features of the s-channel due to the high mass $Z^{'}$ resonance is exploited by reconstructing the invariant mass of the di-jet system. To improve mass resolution, we included information from the missing momentum, computing the missing transverse mass, $M_{\text{T}}$, as done in~\cite{Cohen:2015toa}.
A realistic trigger selection is expected to sculpt $M_{\text{T}}$. Therefore, we require $M_{\text{T}} > 1500$~GeV.
A minimal amount of $\cancel{E}_\text{T}$ is required by asking the transverse ratio $R_{\text{T}} = \cancel{E}_{\text{T}}/ M_{\text{T}}$ to be larger than 0.15. This last requirement does not affect the $M_{\text{T}}$ distribution shape. \\
Finally, we require the minimum azimuthal opening $\Delta \phi_{\text{min}}(j_{1,2},\cancel{E}_{\text{T}})$ between either of the two AK8 jets and the $\cancel{E}_{\text{T}}$ to be less than 0.8. This selection allows to suppress $\text{W}/\text{Z} + \text{jets}$ backgrounds, and select events where the missing momentum is aligned to one of the jets as expected for SVJ events, unlike previous multijet+$\cancel{E}_{\text{T}}$ searches~\cite{CMS:2021far,ATLAS:2021kxv}. \\
For selecting SVJ$\ell$ events, we add  some requirements on the leptons. 
The selected electrons must have a transverse momentum well above the CMS ECAL noise and meet current trigger requirements, namely $p_{\text{T}} > 10$~GeV in the barrel ($0<|\eta|<1.44$) and in the end-caps ($1.57 <|\eta|< 2.6$). We require muons with $|\eta| < 2.5$ and $p_{\text{T}}>10$~GeV such that they can be good seeds for the CMS outside-in global muon reconstruction~\cite{Sirunyan_2018}.
Both electrons and muons are required to have a small transverse impact parameter, $|d_0| < 100 \ \mu$m, since SVJ$\ell$ leptons are produced by dark hadrons prompt decays. This requirement allows to further suppress $\text{t} \bar{\text{t}} + \text{jets}$ background as well as the QCD events where leptons are produced by long-lived b-flavored hadrons decays. \\
We use standard isolation to veto all events with more than one isolated lepton ($\text{I} (\mathcal{\ell}) < 0.1$) since the signal is characterized by pairs of non-isolated opposite-charge leptons. Moreover, we require to have at least one pair of opposite-sign non-inter-isolated leptons \mbox{($\text{I}_{\text{int}} (\mathcal{\ell}) > 0.1$)}. This last requirement improves by more than $95 \%$ QCD rejection and more than $75 \%$ $\text{t} \bar{\text{t}} + \text{jets}$ rejection compared to SVJ-tag selections. Also the electro-weak background contribution is heavily suppressed due to the charge and inter-isolation requirements. \\ \\
To estimate the sensitivity of a dedicated LHC search, we perform a bump hunt in the di-jet $M_{\text{T}}$ spectrum~(Figure~\ref{fig:Limits}\textcolor{blue}{.a}).
In addition, the model predicts resonances appearing in the di-lepton spectrum at the GeV scale. However, the low mass di-lepton signal strongly depends on the details of the dark sector. This model-dependent feature can be exploited by applying a selection on the di-lepton invariant mass spectrum, thus enhancing the sensitivity reach of the search. Here we report the expected sensitivity reach of the LHC remaining agnostic on the di-lepton mass range, together with an estimate of the sensitivity improvement brought by an additional selection on non-inter-isolated leptons pairs with opposite charges requiring their invariant mass to be consistent with the $\rho_v$ mass ($12<m_{\ell \ell}<19$~GeV).  
\noindent

\section{Results}

We have estimated the expected exclusion limit at 95~\% confidence level (CL) for $\sigma \times \text{Br}$ for different masses $M_{Z'}$ using the modified frequentist approach $\text{CL}_{\text{s}}$ in the asymptotic approximation~\cite{Junk:1999kv,Read:2002hq,Cowan:2010js}. Systematic uncertainties related to the trigger (2 \%) and luminosity measurement (1.6 \%) have been included in the binned likelihood template fit of the $M_{\text{T}}$ spectrum as nuisance parameters  with a log-normal prior distribution~\cite{CMS:2021dzg}. No uncertainties related to the shape of the $M_{\text{T}}$ distribution have been included. As shown in Figure \ref{fig:Limits}\textcolor{blue}{.b} for a benchmark $r_{\text{inv}}= 0.3$, the LHC with full Run 2 luminosity is expected to claim the evidence (exclusion) of a leptophobic $Z'$ boson with SVJ${\ell}$ signature and SM couplings up to masses of \mbox{$\sim 3.5$ TeV (4.5 TeV)}. We estimate that adding the requirement on the di-lepton invariant mass  allows to reach the $5 \sigma$ discovery of the $Z'$ boson up to $\sim 3.5$ TeV and improve the exclusion up to more than 5~TeV. The sensitivity to the signal diminishes with higher values of $r_{\text{inv}}$ since the missing momentum tends to be less and less aligned with the jet axis. Moreover, as shown in \ref{fig:Limits}\textcolor{blue}{.b}, the sensitivity of the hadronic SVJ search is from  one up to two orders of magnitude worse with respect to a SVJ$\ell$-dedicated search. We have verified that this loss in sensitivity of the hadronic analysis holds for all values of $r_{\text{inv}}$. 

\section{Discussion and conclusions}
This Letter proposed a new search strategy for the discovery of confining hidden sectors giving rise to semi-visible jets with non-isolated prompt lepton pairs. We have introduced a~simplified model based on two messenger bosons separated by a large mass gap to allow leptonic decays of the composite DM. The phase space of the SVJ$\ell$ signal is expected to be loosely constrained by the existing searches. We have shown that the fully hadronic SVJ s-channel search has poor sensitivity to the SVJ$\ell$ signature due to the mini-isolation lepton veto. Remaining agnostic on the shapes and number of di-lepton resonances produced by the dark hadrons decays, we provided expected exclusion limits performing a bump hunt in the di-jet transverse mass. The proposed search can claim the evidence (exclude) the heavier mediator up to masses of 3.5 TeV (4.5 TeV) with \mbox{$\mathcal{L}_{int} = 138 \ {\text{fb}^{-1}}$}. We have estimated that selecting events in the mass range around the expected di-lepton resonance the analysis can achieve the $5 \sigma$ discovery up to masses of 3.5~TeV and improve the exclusion up to more than 5~TeV. The sensitivity to higher values of $r_{\text{inv}}$ can be significantly enhanced by exploiting the resonant shape of the di-lepton spectrum. \\
Further extensions of the current study on SVJ$\ell$ can go in two main directions. On the one hand, from a phenomenological perspective, one can study signatures where dark pseudo-scalars can decay through the $A'$ portal to $\tau$ leptons via a mass insertion. Or, it would be worth to investigate signatures where the dark hadrons have sizable lifetimes. In~this last case, the main background is instrumental. If the displacement is larger than the tracker volume, no track parameterization is available to treat this case with  the current fast simulation tools.  
On the other hand, for the prompt signature, one can develop dedicated taggers for SVJ$\ell$ exploiting the additional substructure information related to the leptons. In this direction, a classifier exploiting both standard jet substructure and other variables capturing mutual relations between the leptons, such as the inter-isolation variable proposed here, is expected to improve  the signal-to-background discrimination and the sensitivity of the analysis drastically. Another possibility is to adopt signal agnostic strategies based on anomaly detection techniques~\cite{Canelli:2021aps} to identify leptons-enriched semi-visible jets. \\
Finally, the model proposed in this Letter allows also for a low mass production mechanism of leptons-enriched semi-visible jets where dark quarks come directly from the $A'$ boson. Current limitations imposed by the trigger selection strategy require the development of dedicated algorithms to probe this production channel. In addition,  the  boosted event topology, resulting from high momentum and  low mass of $A'$, would lead the production of two collimated semi-visible jets, appearing as merged into a single large jet. This topology requires a dedicated reconstruction strategy that is not addressed in this Letter.

\begin{acknowledgements}
 We want to acknowledge S. Knapen, M. Strassler and T. Cohen for useful suggestions and dialogue. We are grateful to H. Beauchesne for remarking relevant aspects for theoretical consistency of the model. We also thank M. Selvaggi for useful support with the fast simulation framework. We thank F. Eble and J. Niedziela for supporting  the framework used in the analysis. C. Cazzaniga and A. de Cosa are supported by the Swiss National Science Fundation (SNFS) under the SNSF Eccellenza program.
\end{acknowledgements}


\bibliographystyle{spphys}       
\bibliography{bibl}   

%
%

\end{document}